\documentclass[twocolumn]{aastex63}

\renewcommand{\today}{2020 January 31} % Force draft date

\usepackage{microtype}
\usepackage{amsmath}
\usepackage{mathtools} % Needed for dcases environment

\let\tablenum\relax % Avoid conflict with AASTeX
\usepackage[separate-uncertainty, retain-explicit-plus]{siunitx}
\usepackage{graphicx}

\DeclareMathOperator{\Tr}{Tr}

\usepackage{xcolor}
\definecolor{tab:blue}{RGB}{31, 119, 180}
\definecolor{tab:orange}{RGB}{255, 127, 14}
\definecolor{tab:red}{RGB}{214, 39, 40}
\definecolor{plotlightgray}{gray}{0.8} % This is darker than what's used in the plot to try to compensate for the small indicator size
\definecolor{xkcd:darkpurple}{RGB}{53, 6, 62}

\usepackage{soul}
\usepackage[outline]{contour}
\newcommand \colorindicator[2]{%
  \begingroup%
  \setul{0.25ex}{0.4ex}%
  \contourlength{0.2ex}%
  \setulcolor{#1}%
  \ul{{\phantom{#2}}}\llap{\contour{white}{#2}}~\textcolor{#1}{\tiny{$\blacksquare$}}%
  \endgroup%
}

\usepackage{CJKutf8}
\newcommand{\cntext}[1]{\begin{CJK*}{UTF8}{gbsn}#1\end{CJK*}}
\usepackage{devanagari}

\begin{document}

\title{Two-year Cosmology Large Angular Scale Surveyor (CLASS) Observations:\\A First Detection of Atmospheric Circular Polarization at Q~Band}
\author[0000-0002-4436-4215]{Matthew~A. Petroff}
\author[0000-0001-6976-180X]{Joseph~R. Eimer}
\affiliation{Department of Physics \& Astronomy, Johns Hopkins University, Baltimore, Maryland 21218, USA}
\author[0000-0003-1248-9563]{Kathleen Harrington}
\affiliation{Department of Physics, University of Michigan, Ann Arbor, Michigan 48109, USA}
\affiliation{Department of Physics \& Astronomy, Johns Hopkins University, Baltimore, Maryland 21218, USA}
\author[0000-0001-7941-9602]{Aamir Ali}
\affiliation{Department of Physics, University of California, Berkeley, California 94720, USA}
\affiliation{Department of Physics \& Astronomy, Johns Hopkins University, Baltimore, Maryland 21218, USA}
\author[0000-0002-8412-630X]{John~W. Appel}
\author[0000-0001-8839-7206]{Charles~L. Bennett}
\author{Michael~K. Brewer}
\affiliation{Department of Physics \& Astronomy, Johns Hopkins University, Baltimore, Maryland 21218, USA}
\author[0000-0001-8468-9391]{Ricardo Bustos}
\affiliation{Facultad de Ingenier\'ia, Universidad Cat\'olica de la Santísima Concepci\'on, Concepci\'on, Chile}
\author[0000-0003-1127-0965]{Manwei Chan}
\affiliation{Department of Physics \& Astronomy, Johns Hopkins University, Baltimore, Maryland 21218, USA}
\author[0000-0003-0016-0533]{David T. Chuss}
\affiliation{Department of Physics, Villanova University, Villanova, Pennsylvania 19085, USA}
\author{Joseph Cleary}
\author[0000-0002-0552-3754]{Jullianna Denes Couto}
\author[0000-0002-1708-5464]{Sumit Dahal ({\dn \7{s}Emt dAhAl})}
\affiliation{Department of Physics \& Astronomy, Johns Hopkins University, Baltimore, Maryland 21218, USA}
\author{Rolando D\"unner}
\affiliation{Instituto de Astrof\'isica, Facultad de F\'isica, Pontificia Universidad Cat\'olica de Chile, 7820436 Macul, Santiago, Chile}
\affiliation{Centro de Astro-Ingenier\'ia, Facultad de F\'isica, Pontificia Universidad Cat\'olica de Chile, 7820436 Macul, Santiago, Chile}
\author[0000-0002-4782-3851]{Thomas Essinger-Hileman}
\affiliation{Code 665, NASA Goddard Space Flight Center, Greenbelt, Maryland 20771, USA}
\author[0000-0002-2061-0063]{Pedro {Flux\'a~Rojas}}
\affiliation{Instituto de Astrof\'isica, Facultad de F\'isica, Pontificia Universidad Cat\'olica de Chile, 7820436 Macul, Santiago, Chile}
\affiliation{Centro de Astro-Ingenier\'ia, Facultad de F\'isica, Pontificia Universidad Cat\'olica de Chile, 7820436 Macul, Santiago, Chile}
\author{Dominik Gothe}
\author[0000-0001-7466-0317]{Jeffrey Iuliano}
\author[0000-0003-4496-6520]{Tobias~A. Marriage}
\affiliation{Department of Physics \& Astronomy, Johns Hopkins University, Baltimore, Maryland 21218, USA}
\author{Nathan J. Miller}
\affiliation{Department of Physics \& Astronomy, Johns Hopkins University, Baltimore, Maryland 21218, USA}
\affiliation{Code 665, NASA Goddard Space Flight Center, Greenbelt, Maryland 20771, USA}
\author[0000-0002-5247-2523]{Carolina N\'u\~nez}
\author[0000-0002-0024-2662]{Ivan L. Padilla}
\affiliation{Department of Physics \& Astronomy, Johns Hopkins University, Baltimore, Maryland 21218, USA}
\author[0000-0002-8224-859X]{Lucas Parker}
\affiliation{Space and Remote Sensing, MS B244, Los Alamos National Laboratory, Los Alamos, New Mexico 87544, USA}
\author[0000-0001-5704-271X]{Rodrigo Reeves}
\affiliation{CePIA, Departamento de Astronom\'ia, Universidad de Concepci\'on, Concepci\'on, Chile}
\author[0000-0003-4189-0700]{Karwan Rostem}
\affiliation{Code 665, NASA Goddard Space Flight Center, Greenbelt, Maryland 20771, USA}
\author{Deniz Augusto Nunes Valle}
\author[0000-0002-5437-6121]{Duncan~J. Watts}
\author[0000-0003-3017-3474]{Janet~L. Weiland}
\affiliation{Department of Physics \& Astronomy, Johns Hopkins University, Baltimore, Maryland 21218, USA}
\author[0000-0002-7567-4451]{Edward J.~Wollack}
\affiliation{Code 665, NASA Goddard Space Flight Center, Greenbelt, Maryland 20771, USA}
\author[0000-0001-5112-2567]{Zhilei Xu (\cntext{徐智磊}\!\!)}
\affiliation{Department of Physics \& Astronomy, University of Pennsylvania, Philadelphia, Pennsylvania 19104, USA}
\affiliation{Department of Physics \& Astronomy, Johns Hopkins University, Baltimore, Maryland 21218, USA}

\shorttitle{Atmospheric Circular Polarization at Q~Band}
\shortauthors{Petroff et al.}

\correspondingauthor{Matthew A. Petroff}
\email{petroff@jhu.edu}

\keywords{\href{http://astrothesaurus.org/uat/322}{Cosmic microwave background radiation (322)}; \href{http://astrothesaurus.org/uat/1146}{Observational cosmology (1146)}; \href{http://astrothesaurus.org/uat/1277}{Polarimeters (1277)}; \href{http://astrothesaurus.org/uat/799}{Astronomical instrumentation (799)}; \href{http://astrothesaurus.org/uat/113}{Atmospheric effects (113)}}

\begin{abstract}
\noindent The Earth's magnetic field induces Zeeman splitting of the magnetic dipole transitions of molecular oxygen in the atmosphere, which produces polarized emission in the millimeter-wave regime. This polarized emission is primarily circularly polarized and manifests as a foreground with a dipole-shaped sky pattern for polarization-sensitive ground-based cosmic microwave background experiments, such as the Cosmology Large Angular Scale Surveyor (CLASS), which is capable of measuring large angular scale circular polarization. Using atmospheric emission theory and radiative transfer formalisms, we model the expected amplitude and spatial distribution of this signal and evaluate the model for the CLASS observing site in the Atacama Desert of northern Chile. Then, using two years of observations at \SIrange{32.3}{43.7}{\giga\hertz} from the CLASS Q-band telescope, we present a detection of this signal and compare the observed signal to that predicted by the model. We recover an angle between magnetic north and true north of \SI{-5.5 \pm 0.6}{\degree}, which is consistent with the expectation of \SI{-5.9}{\degree} for the CLASS observing site. When comparing dipole sky patterns fit to both simulated and data-derived sky maps, the dipole directions match to within a degree, and the measured amplitudes match to within ${\sim}20\%$.
\end{abstract}

\section{Introduction}

In the presence of Earth's magnetic field, molecular oxygen in the atmosphere experiences Zeeman splitting of its magnetic dipole transitions. This produces polarized emission in the millimeter-wave regime, primarily circular polarization, which manifests as a foreground for polarization-sensitive ground-based cosmic microwave background (CMB) experiments such as the Cosmology Large Angular Scale Surveyor (CLASS) \citep{Eimer2012, EssingerHileman2014, Harrington2016}. The effect of this foreground has previously been discussed qualitatively by \citet{Keating1998} and quantified by \citet{Hanany2003} and \citet{Spinelli2011}. However, previous attempts to observe this foreground, such as by \textsc{Mipol} \citep{Mainini2013}, have not been successful in detecting it.

Molecular oxygen has strong emission lines in the \SIrange{50}{70}{\giga\hertz} range, as well as a line at \SI{118.8}{\giga\hertz}, which are in the frequency range of interest for CMB observations. There are also water vapor and ozone emission lines near this frequency range, but unlike diatomic oxygen, these molecules do not experience Zeeman splitting and thus do not produce polarized emission \citep{Liebe1981}. Zeeman splitting of molecular oxygen is generally considered in the context of remote sensing of the temperature of the mesosphere, where individual Zeeman-split emission lines can be resolved; pressure broadening obscures individual lines at lower altitudes \citep{Meeks1963}.

Modeling of the polarized emission of Zeeman-split oxygen began with the seminal works of \citet{Lenoir1967, Lenoir1968}, with further development of atmospheric emission models by \citet{Liebe1981, Liebe1989}, \citet{Liebe1992, Liebe1993}, \citet{Rosenkranz1988}, and others. Near the emission lines, both linearly and circularly polarized emission have been detected from orbit \citep{Schwartz2006, Kunkee2008}. Ground-based detections of linear polarization have also been made at the emission lines, at \SI{234}{\giga\hertz} \citep{Pardo1995} and \SI{53}{\giga\hertz} \citep{NavasGuzmn2015}. However, atmospheric remote sensing instruments generally do not observe polarization at frequencies far from the emission lines.

As the circular polarization predicted by standard cosmological models is many orders of magnitude smaller than that of the linearly polarized signal \citep{Inomata2019}, CMB experiments are built to be primarily sensitive to linear polarization, which is itself an extremely faint signal \citep{Hu1997}. For ground-based experiments, the linearly polarized signal from Zeeman-split molecular oxygen is expected to be on the \si{\nano\kelvin} level, roughly four orders of magnitude weaker than the corresponding circularly polarized signal \citep{Hanany2003}. As this is much fainter than can be detected with current CMB instruments and as the circularly polarized component is much stronger, atmospheric Zeeman emission is primarily considered a CMB foreground at the largest angular scales ($\ell \gtrsim 2$), fixed in topocentric coordinates, due to potential circular-to-linear polarization leakage as a result of instrument non-idealities \citep{ODea2007}. For linear polarization, emission and scattering by ice crystal clouds in the upper troposphere is a larger atmospheric contaminant \citep{Pietranera2007, Takakura2019}.

For large angular scales at Q~band (\SIrange{\sim30}{\sim50}{\giga\hertz}), diffuse Galactic synchrotron emission is expected to be the largest contributor to extraterrestrial circular polarization, with circular polarization due to Faraday conversion of linear polarization induced by Population III stars exceeding this contribution at smaller angular scales \citep{King2016}. Another possible source of Faraday conversion is via galaxy cluster magnetic fields, at primarily small angular scales \citep{Cooray2003}. Beyond synchrotron emission and Faraday conversion, additional potential sources of circular polarization include scattering by the cosmic neutrino background \citep{Mohammadi2014}, primordial magnetic fields \citep{Giovannini2009}, photon--photon interactions in neutral hydrogen \citep{Sawyer2015}, and cosmic birefringence via coupling of the Chern--Simons term \citep{Carroll1990}, as well as postulated new physics \citep{Zarei2010, Tizchang2016}. As these predicted signals are at most on the \si{\nano\kelvin} level, they are well below current detection thresholds.

The most stringent previously published upper limit on CMB circular polarization was set by the balloon-borne \textsc{Spider} instrument, utilizing non-idealities in its half-wave plate polarization modulators at frequencies near \SI{95}{\giga\hertz} and \SI{150}{\giga\hertz} \citep{Nagy2017}. As \textsc{Spider} observed from the stratosphere, above much of the atmosphere and thus much of the atmospheric emission, it was not sensitive to atmospheric circularly polarized emission. However, for ground-based experiments to significantly improve on this limit, circularly polarized atmospheric emission must first be detected and subtracted. As with \textsc{Mipol} \citep{Mainini2013}, previous ground-based measurements did not have the requisite sensitivity to detect circularly polarized atmospheric emission \citep{Lubin1983, Partridge1988}. An improved upper limit on extraterrestrial circular polarization utilizing CLASS observations is presented in a companion paper, \citet{Padilla2019}.

Through its use of Variable-delay Polarization Modulators (VPMs) \citep{Chuss2012, Harrington2018}, CLASS is uniquely capable of measuring large angular scale circular polarization. CLASS currently observes the microwave sky in frequency bands centered near \SIlist{40;90;150;220}{\giga\hertz}, using three telescope receivers; a fourth receiver will be deployed in the future. The present analysis focuses on the first two years of observations from the Q-band telescope, which is centered near \SI{40}{\giga\hertz}. CLASS is designed to map the polarization of the CMB at large angular scales over 75\% of the sky to detect or place an upper limit on the B-mode signal of primordial gravitational waves and to measure the optical depth due to reionization, $\tau$, to near the cosmic variance limit \citep{Watts2015, Watts2018}.

VPMs utilize a movable mirror placed behind a linearly polarizing array of parallel wires to induce a varying phase delay between polarization states perpendicular to and parallel to the direction of the array wires. When combined with detectors sensitive to linear polarization, modulating the mirror position, and thus the distance between the mirror and the wire array, results in the modulation of one linear polarization state, instrument Stokes $U$ in the case of CLASS, into circular polarization, Stokes $V$, and vice versa. This modulation increases polarization measurement stability by utilizing phase-sensitive detection and allows for the measurement of large angular scale modes on the sky. As Stokes $U$ and $V$ are modulated instead of Stokes $Q$ and $U$ as is the case of half-wave plate modulators more commonly used in CMB instruments \citep{Kusaka2018}, CLASS has significant capability to measure circular polarization (K. Harrington et al. 2020, in preparation). Furthermore, its $V$ detection capability is a more direct measurement than sensitivity obtained via half-wave plate non-idealities, which are highly frequency dependent in a poorly constrained manner \citep{Nagy2017}.

The remainder of this paper is organized as follows. In Section \ref{sec:theory}, we present the theory behind polarized Zeeman emission of molecular oxygen in the atmosphere that is used to simulate the expected signal. Next, in Section \ref{sec:sims}, we present the results of these simulations for the CLASS observing site in the Atacama Desert of northern Chile. Then, we compare the simulation results to data from the first era of observations of the CLASS Q-band receiver in Section \ref{sec:observations}. Finally, we conclude in Section \ref{sec:conclusion}.

\section{Atmospheric emission theory}
\label{sec:theory}

In its electronic ground state, molecular oxygen has a spin quantum number $S = 1$ due to the unpaired spins of two electrons, resulting in a magnetic dipole moment. This magnetic dipole moment results in transitions between rotational states of the molecule's electronic and vibrational ground states with millimeter-wave emission. The spin quantum number couples with the total rotational angular momentum quantum number $N$, which must be odd due to the exclusion principle, to yield the rotational quantum number $J$. This results in three possible values for $J$ per $N$, $J=N,N\pm1$. Selection rules allow for an $N^+$ transition, $(J=N) \to (J=N+1)$, and an $N^-$ transition, $(J=N) \to (J=N-1)$. In the absence of an external magnetic field, this emission is unpolarized, but a non-zero external magnetic field induces Zeeman splitting, which produces polarized emission \citep{Berestetskii1982}. An external magnetic field splits the transition lines due to a given $J$ into $2J+1$ lines corresponding to the magnetic quantum number $M$, where $-J \leq M \leq J$; $M$ expresses the projection of the molecular magnetic moment on the external magnetic field vector.

\subsection{Layer Attenuation}

\subsubsection{Molecular Oxygen}

In a coherency matrix formalism \citep{Lenoir1967}, the attenuation of a given atmosphere layer due to resonances in molecular oxygen can be defined as
\begin{equation}
\boldsymbol G_Z(\nu) = \frac{1}{2}\sum_i S_i \sum_{\Delta M = -1}^1 \boldsymbol\rho \sum_{M = -N}^N \xi(N, M) F(\nu, \nu_k),
\end{equation}
where $S_i$~(\si{\giga\hertz\per\kilo\meter}) is the intensity of the unsplit line, $\boldsymbol\rho$ is the transition matrix, $\xi$ is the intensity of the Zeeman component relative to the unsplit line, $F$~(\si{\giga\hertz}) is the line profile, $\nu$~(\si{\giga\hertz}) is the frequency of evaluation, and $\nu_k$~(\si{\giga\hertz}) is the frequency of the Zeeman line. Note that this is field attenuation, which is a factor of two smaller than power attenuation, in units of \si{\neper\per\kilo\meter}. The outer summation is performed over the first thirty-eight resonance lines, starting at $N^\pm=1$ and ending at $N^\pm=37$. Higher order resonance lines are excluded due to lack of available line broadening and mixing data; as these lines are much weaker, the effect of excluding them is minimal.

The intensity of the unsplit line (\si{\giga\hertz\per\kilo\meter}) is
\begin{equation}\label{eq:lineint}
S_i = S_{296}  \frac{n p_\mathrm{air} P}{RT} \left(\frac{T_{296}}{T}\right)^{2.5}\exp\left[\frac{E''}{T_{296} k_B}\left(1-\frac{T_{296}}{T}\right)\right],
\end{equation}
where $S_{296}$~(\si{\mega\hertz\square\meter\per\mole}) is the fiducial intensity at \SI{296}{\kelvin}, $n \approx 0.2095$ is the volume fraction of molecular oxygen in the atmosphere \citep{Machta1970},\footnote{This decreases at very high altitudes due to photodissociation \citep{Penndorf1949}, but, as will be shown later, the relevant signal is primarily from the lower atmosphere.} $p_\mathrm{air} = 1 - p_w$ is the fractional partial pressure of dry air, $P$ (\si{\hecto\pascal}) is the pressure of the atmosphere layer, $R$~(\si{\joule\per\mole\per\kelvin}) is the molar gas constant, $T$~(\si{\kelvin}) is the physical temperature of the atmosphere layer, $T_{296} = \text{\SI{296}{\kelvin}}$ is a reference temperature, $E''$ (\si{\joule}) is the lower-state energy of the transition, and $k_B$ is the Boltzmann constant~(\si{\joule\per\kelvin}). $S_{296}$ and $E''$ values are from \textsc{Hitran} \citep{Gordon2017}. As per \citet{Liebe1993}, the fractional partial pressure of water vapor is
\begin{equation}
p_w = 2.408\cdot 10^{11}\frac{u}{P}\left(\frac{T_{300}}{T}\right)^5\exp\left[-22.644 \left(\frac{T_{300}}{T}\right)\right],
\end{equation}
where $u$ is the fractional relative humidity and $T_{300} = \text{\SI{300}{\kelvin}}$ is a reference temperature.

For $\Delta M = 0$,
\begin{equation}
\boldsymbol\rho_\pi = \begin{pmatrix}
0 & 0 \\
0 & \sin^2\theta
\end{pmatrix} = \frac{\sin^2\theta}{2}\boldsymbol\sigma_I - \frac{\sin^2\theta}{2}\boldsymbol\sigma_Q,
\end{equation}
and for $\Delta M = \pm 1$,
\begin{equation}
\begin{split}
\boldsymbol\rho_{\sigma\pm} &= \begin{pmatrix}
1 & \mp i \cos\theta \\
\pm i \cos\theta & \cos^2\theta
\end{pmatrix} \\&= \frac{1+\cos^2\theta}{2}\boldsymbol\sigma_I + \frac{1-\cos^2\theta}{2}\boldsymbol\sigma_Q \pm\cos\theta\boldsymbol\sigma_V,
\end{split}
\end{equation}
where $\theta$ is the angle between the line of sight and the geomagnetic field vector and
\begin{equation*}
\boldsymbol\sigma_I = \begin{pmatrix}
1 & 0 \\
0 & 1
\end{pmatrix},\quad\boldsymbol\sigma_Q = \begin{pmatrix}
1 & 0 \\
0 & -1
\end{pmatrix},\quad\boldsymbol\sigma_V = \begin{pmatrix}
0 & -i \\
i & 0
\end{pmatrix}.
\end{equation*}
Thus, the amplitude of circular polarization is maximized when the line of sight is aligned with the direction of the magnetic field, and the linear polarization is maximized when the two vectors are perpendicular.

While the Zeeman effect in molecular oxygen can be approximated to reasonable accuracy by Hund's case (b) \citep{Townes1955}, we use the more exact calculations laid out in \citet{Larsson2019}. In this formalism, the frequencies of the Zeeman lines are given by $\nu_k = \nu_i + \delta_{\Delta M}$, where $\nu_i$ (\si{\giga\hertz}) [from \textsc{Hitran} \citep{Gordon2017}] is the frequency of the unsplit line and
\begin{equation}
\delta_{\Delta M} = -\frac{\mu_B B}{h}\left[ g_{N=J} M + g_{N=J\pm 1} (M + \Delta M) \right]\cdot 10^{-9}
\end{equation}
is the Zeeman frequency shift in \si{\giga\hertz}. Here, $B$~(\si{\tesla}) is the magnetic field strength, $\mu_B$~(\si{\joule\per\tesla}) is the Bohr magneton, $h$~(\si{\joule\second}) is the Planck constant, and $g_{N=J}$ and $g_{N=J\pm 1}$ are the numerical Zeeman coefficients corresponding to the given $N^\pm$ transition from Table 2 of \citet{Larsson2019}. The values of the relative intensity factor $\xi(N, M)$ are shown in Table \ref{tab:zeemanint}; note that these normalize to two in the absence of a magnetic field, but when combined with $\boldsymbol\rho$, the combination yields the identity matrix.

\begin{deluxetable*}{lcc}
\tablecaption{Relative intensity factor $\xi(N, M)$\label{tab:zeemanint}}
\tablehead{
\colhead{} &
\colhead{$N^+$ line} &
\colhead{$N^-$ line}
}
\startdata \\[-1em]
$\Delta M = 0$ & $\displaystyle\frac{3[(N+1)^2-M^2]}{(N+1)(2N+1)(2N+3)}$ & $\displaystyle\frac{3(N+1)(N^2-M^2)}{N(2N+1)(2N^2+N-1)}$ \\[1em]
$\Delta M = \pm 1$ & $\displaystyle\frac{3 (N \pm M+1)(N \pm M+2)}{4(N+1)(2N+1)(2N+3)}$ & $\displaystyle\frac{3(N+1)(N \pm M)(N \pm M-1)}{4N(2N+1)(2N^2+N-1)}$ \\[1em]
\enddata
\tablecomments{For Zeeman components of O\textsubscript{2} lines, where $\pm M \leq N$ \citep{Liebe1981}.}
\end{deluxetable*}

Following \citet{Larsson2014} and \citet{Melsheimer2005} the line profile is defined as
\begin{equation}
\begin{split}
F(\nu, \nu_k) &= \left(\frac{\nu}{\nu_k}\right)^2\frac{1}{\Delta \nu_D\sqrt{\pi}}\left[(1 + g_l P^2 - i y_l P) w(z_-)\right. \\
&\quad\left.{}+ (1 + g_l P^2 + i y_l P) w(z_+)\right],
\end{split}
\end{equation}
where the Faddeeva function (for $\operatorname{Im}(z) > 0$) \citep{Faddeyeva1961} is,
\begin{equation}
w(z) = \frac{i}{\pi} \int_{-\infty}^\infty \frac{e^{-t^2}}{z-t}dt,
\end{equation}
with
\begin{equation}
z_\pm = \frac{\nu \pm \nu_k \pm \delta \nu_l P^2 + i\Delta \nu_p}{\Delta \nu_D}.
\end{equation}
Here, $g_l$ is the second order line shape correction, $y_l$ is the first order phase correction, and $\delta \nu_l$ is the second order frequency correction. These are defined with temperature dependence by
\begin{equation}
Z_l(T) = \left[Z_l^0 + Z_l^1\left(\frac{T_{300}}{T}-1\right)\right]\left(\frac{T_{300}}{T}\right)^{x_Z},
\end{equation}
where $Z_l$ is the first or second order coefficient $y_l$, $g_l$, or $\delta\nu_l$ and $y_l^{0,1}$ (\si{\per\hecto\pascal}), $g_l^{0,1}$ (\si{\per\square\hecto\pascal}), and $\delta\nu_l^{0,1}$ (\si{\giga\hertz\per\square\hecto\pascal}) are all from Table 1 of \citet{Makarov2011}; the exponent $x_Z$ is 0.8 for $y_l$ and 1.6 for $g_l$ and $\delta\nu_l$. The pressure broadening half width is
\begin{equation}
\Delta \nu_p = \gamma_\mathrm{air} p_\mathrm{air}P\left(\frac{T_{296}}{T}\right)^{x_\nu} + \gamma_w p_w P\left(\frac{T_{296}}{T}\right)^{x_\nu}
\end{equation}
for
\begin{equation}
\gamma_{\mathrm{air},w}(N) = A_\gamma + \frac{B_\gamma}{1 + c_1N + c_2N^2+c_3N^4},
\end{equation}
where $A_\gamma$ (\si{\giga\hertz\per\hecto\pascal}), $B_\gamma$ (\si{\giga\hertz\per\hecto\pascal}), $c_1$, $c_2$, and $c_3$ are parameters from column four of Table 3 of \citet{Koshelev2016} for $\gamma_\mathrm{air}$ and from column two of Table 2 of \citet{Koshelev2015} for $\gamma_w$; the exponent $x_\nu = 0.75412$ is from Table 1 of \citet{Koshelev2016}.
Also used is the $1/e$ Doppler half width \citep{Herbert1974, Varghese1984},
\begin{equation}
\Delta\nu_D = \frac{\nu_k}{c}\sqrt{\frac{2 k_B T}{M_{O_2}}}
\end{equation}
in \si{\giga\hertz}, where $c$ (\si{\meter\per\second}) is the speed of light and $M_{O_2}$ (\si{\kilo\gram\per\mole}) is the molar mass of molecular oxygen [from \textsc{Hitran} \citep{Gordon2017}]. The $1/(\Delta \nu_D\sqrt{\pi})$ factor is a normalization \citep{Armstrong1967}. At high pressures, the Faddeeva function simplifies to a Lorentz shape function, so the combination of the two Faddeeva functions reduces to a Van~Vleck--Weisskopf \citep{VanVleck1945} line shape. At low pressures, when $P \to 0$, the line mixing and pressure correction effects are eliminated, so the function behaves similar to a Voigt line shape function, although with slightly reduced amplitude on the line wings. As will later be shown, the high pressure case is more important, so the line shape function was chosen such that it is most accurate in that regime.

\subsubsection{Dry Air and Water Vapor}

While the Zeeman effect is the sole significant non-transient source of polarized atmospheric emission in the millimeter spectrum, there are other unpolarized sources, quantified by a dry air continuum and water vapor contributions. Following \citet{Tretyakov2016}, the dry air and water vapor non-resonant continua term is
\begin{equation}
\begin{split}
\alpha_c(\nu, T) &= \left[C_w^0 \left(\frac{T_{300}}{T}\right)^{x_w}p_w^2 + C_\mathrm{air}^0 \left(\frac{T_{300}}{T}\right)^{x_\mathrm{air}}p_\mathrm{air}p_w \right. \\
&\quad \left.{} + C_\mathrm{dry}^0 \left(\frac{T_{300}}{T}\right)^{x_\mathrm{dry}}p_\mathrm{air}^2\right]P^2\nu^2,
\end{split}
\end{equation}
where $C_w^0$, $C_\mathrm{air}^0$, $C_\mathrm{dry}^0$ (\si{\neper\per\kilo\meter\per\square\hecto\pascal\per\square\giga\hertz}) and $x_w$, $x_\mathrm{air}$, $x_\mathrm{dry}$ are numerical coefficients from from Table 5 of \citet{Tretyakov2016}. Following \citet{Rosenkranz1998}, the water vapor contribution is
\begin{equation}
\alpha_w(\nu, T) = \sum_j S_j(T) \left[f_j(\nu) + f_j(-\nu)\right]
\end{equation}
summed over the water vapor lines shown in Table 1 of \citet{Rosenkranz1998}, where the line profile is defined as
\begin{equation}
\begin{split}
&f_j(\nu) = \\ &\begin{dcases}
\frac{\nu^2 \gamma_j}{\pi \nu_j^2}\left[ \frac{1}{\left(\nu - \nu_j\right)^2 + \gamma_j^2} - \frac{1}{\nu_c^2 + \gamma_j^2} \right] & |v-v_j| < v_c\\
0 & |v-v_j| \geq v_c
\end{dcases}
\end{split}
\end{equation}
with
\begin{equation}
\gamma_i = w_s p_w P \left(\frac{T_{300}}{T}\right)^{x_s} + w_f p_\mathrm{air} P \left(\frac{T_{300}}{T}\right)^{x_f},
\end{equation}
where $\nu_c = \text{\SI{750}{\giga\hertz}}$ is a cutoff frequency and $w_s$, $w_f$ (\si{\giga\hertz\per\kilo\meter}) and $x_s$, $x_f$ are numerical coefficients from Table 1 of \citet{Rosenkranz1998}. This formulation resembles the Van~Vleck--Weisskopf line shape but includes a high frequency cutoff, as proposed by \citet{Clough1989}. The line intensities, $S_j$ (\si{\giga\hertz\per\kilo\meter}), follow equation (\ref{eq:lineint}) but with $np_\mathrm{air}$ replaced with $p_w$; $S_{296}$ (\si{\mega\hertz\square\meter\per\mole}) and $E''$ (\si{\joule}) values are again from \textsc{Hitran} \citep{Gordon2017}, along with the line frequencies, $\nu_j$ (\si{\giga\hertz}). The full coherency matrix (\si{\neper\per\kilo\meter}) is then
\begin{equation}\label{eq:coherencymatrix}
\boldsymbol G(\nu) = \boldsymbol G_Z(\nu) + \frac{\alpha_c}{2}\boldsymbol I + \frac{\alpha_w}{2}\boldsymbol I,
\end{equation}
where $\boldsymbol I$ is the identity matrix.

\subsection{Radiative Transfer}

As we are interested in polarization, a tensor radiative transfer approach is required to model how radiation propagates through the atmosphere. Using a plane-parallel approximation, the atmosphere is divided into layers of thickness $\Delta z = \sec\phi\cdot\text{\SI{0.2}{\kilo\meter}}$ starting from ground level, \SI{5.2}{\kilo\meter} in the case of CLASS, and ending at \SI{100}{\kilo\meter}; $\phi$ denotes the zenith angle. Using the approach described in \citet{Lenoir1967, Lenoir1968} and including phase as per \citet{Rosenkranz1988}, the brightness temperature coherency matrix \citep{NIST2008} of a given atmosphere layer is defined as
\begin{equation}\label{eq:radiativetransfer}
\begin{split}
\boldsymbol T_B(z) &= e^{-\boldsymbol G\Delta z} \boldsymbol T_B(z_0) e^{-\boldsymbol G^\dagger\Delta z} \\
&\quad + T(z)\left( \boldsymbol I - e^{-\boldsymbol G\Delta z}e^{-\boldsymbol G^\dagger\Delta z}\right)
\end{split}
\end{equation}
using matrix exponentiation, where $\boldsymbol T_B(z_0)$ is the brightness temperature coherency matrix of the atmosphere layer before it, $T(z)$ (\si{\kelvin}) is the physical temperature of the atmosphere layer, and $\dagger$ represents the conjugate transpose operation. Since we are observing from the ground, unlike the satellite observations described in \citet{Lenoir1968}, we start at \SI{100}{\kilo\meter} with $\boldsymbol T_B(z_0)= T_\textrm{CMB}^\textrm{RJ} \boldsymbol I$ and calculate the propagation downward layer-by-layer until the ground, where $T_\textrm{CMB}^\textrm{RJ}$ is the CMB monopole \citep{Fixsen2009} brightness temperature at the observing frequency. Note that as the brightness temperature coherency matrix contains matrix elements defined in terms of the brightness temperature of a single polarization of radiation,\footnote{See the footnote of Section 1.2.2 of \citet{Janssen1993} for a discussion of differing brightness temperature definitions.} there is an extra factor of one half in the conversion to Stokes parameters when compared to a standard coherency matrix \citep{Born1959}, e.g., $I=\frac{1}{2}\Tr(\boldsymbol T_B)$.

\subsection{Atmosphere and Magnetic Field Properties}

In order to perform the aforementioned calculations, data are needed for the atmospheric temperature and pressure profiles, as well as the geomagnetic field vector direction and magnitude at the location in question. The \textsc{Nrlmsise-00} atmosphere model \citep{Picone2002}, averaged over the full year, was used to calculate the temperature and pressure profiles used; for the lower atmosphere, this model was compared to data acquired via radiosonde launches from the Chajnantor Plateau, which is adjacent to the CLASS observing site, during the \textsc{Alma} site characterization campaign\footnote{Data retrieved from \url{http://legacy.nrao.edu/alma/site/Chajnantor/instruments/radiosonde/}.} and was found to be in good agreement. The magnetic field was calculated at ground level using the National Oceanic and Atmospheric Administration's Enhanced Magnetic Model 2017 (EMM2017) \citep{EMM2017}. While the magnetic field strength does decrease with altitude, this effect is minor, so it was omitted from the calculations.

\subsection{Primary Source of Polarized Emission}

\begin{figure}
\includegraphics[width=\columnwidth]{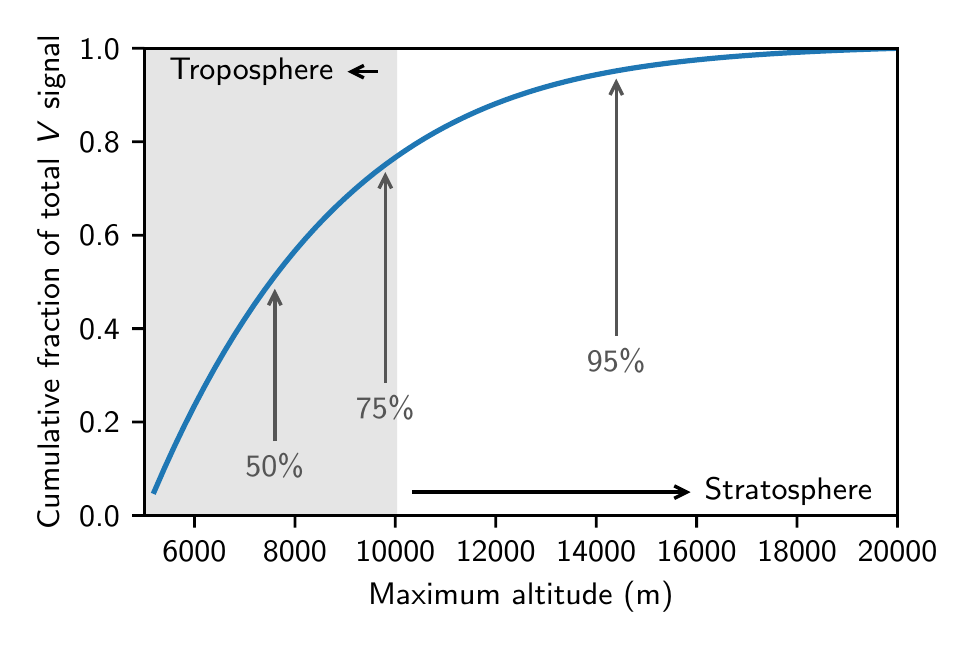}
\caption{Altitude dependence of Zeeman emission, comparing the total Stokes $V$ signal observed for various simulation altitude cutoffs to the fiducial case of \SI{100}{\kilo\meter} maximum altitude. The simulated observations are to the north with a \SI{45}{\degree} zenith angle, starting from \SI{5200}{\meter} at the CLASS observing site for the CLASS Q-band telescope. The light gray background denotes the troposphere, while the white background denotes the stratosphere.}
\label{fig:altitude-effect}
\end{figure}

Zeeman emission from atmospheric molecular oxygen is generally considered to be a mesospheric effect, since individual emission lines can be discerned at these altitudes due to the lack of pressure broadening. However, we are interested in polarized emission at frequencies far from the resonance lines, where the discernibility of individual emission lines is not relevant. To determine which altitude region of the atmosphere is primarily responsible for the polarized emission, we simulated a series of observations to the north with a \SI{45}{\degree} zenith angle from the CLASS observing site in the frequency band of the CLASS Q-band telescope; the properties of both the site and telescope are further detailed in the next section. Instead of starting at \SI{100}{\kilo\meter} altitude to calculate the layer-by-layer propagation downward, the starting altitude is reduced incrementally and compared to the \SI{100}{\kilo\meter} fiducial case, the results of which are shown in Figure \ref{fig:altitude-effect}. This series of calculations shows that the lower atmosphere is the primary contributor to the polarized Zeeman emission far from the resonance lines, with three-quarters of the signal contributed by the troposphere. At lower altitudes, the atmospheric pressure is higher, so there are more oxygen molecules per unit volume, which leads to more emission. Thus, while the mesospheric emission is most significant when one wishes to resolve individual Zeeman emission lines, tropospheric emission is most significant far from the resonance lines.

\section{Simulation results}
\label{sec:sims}

The CLASS observing site is located at \SI{5.2}{\kilo\meter} elevation in the Atacama Desert of northern Chile, with coordinates \SI{22.95975}{\degree}\,S, \SI{67.78726}{\degree}\,W. For this location, at ground level, the EMM2017 magnetic field model was evaluated for 2017 January 1, a date near the middle of the observing period that will be described in Section \ref{sec:observations}; this resulted in a magnetic field of \SI{22738}{\nano\tesla} oriented with an azimuth angle of \SI{-5.9}{\degree} and a zenith angle of \SI{68.8}{\degree}. As previously mentioned, the \textsc{Nrlmsise-00} atmosphere model was used to calculate temperature and pressure profiles. To include water vapor effects, $10\%$ relative humidity is assumed, which is a typical value for the CLASS observing site during good weather; this corresponds to \SI{\sim0.6}{\milli\meter} of precipitable water vapor (PWV). As the observations are made away from the water vapor resonance lines, the relative humidity only affects the polarized signal via effecting small changes to the partial pressure of molecular oxygen. Thus, water vapor effects are small for the frequencies of interest, so the exact value is not critical.

\begin{figure}
\includegraphics[width=\columnwidth]{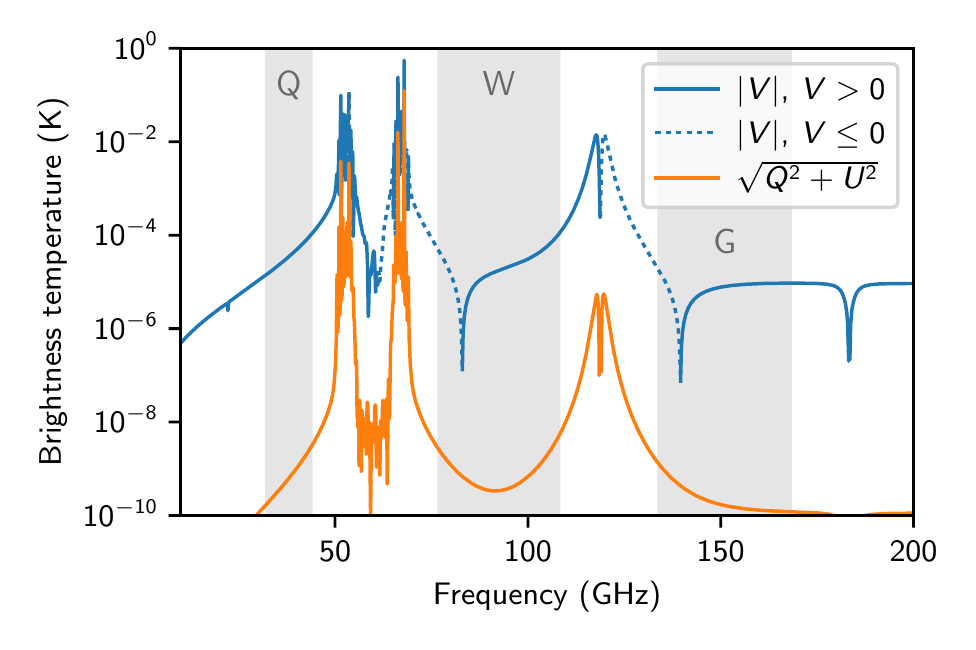}
\caption{Frequency dependence ($\nu$) of polarized atmospheric signal at zenith for the CLASS observing site, both for circular polarization ($|V|$, shown in \colorindicator{tab:blue}{blue}) and linear polarization ($\sqrt{Q^2+U^2}$, shown in \colorindicator{tab:orange}{orange}). The \colorindicator{plotlightgray}{light gray} bands indicate CLASS observing frequencies, with the lowest frequency band corresponding to the Q-band telescope.}
\label{fig:frequency-dependence}
\end{figure}

Using the atmosphere temperature and pressure profiles described above, a full radiative transfer simulation can be performed, as described by equation (\ref{eq:radiativetransfer}), using coherency matrices described by equation (\ref{eq:coherencymatrix}). It is informative to first consider the frequency dependence of the polarized emission, which is shown in Figure \ref{fig:frequency-dependence}. The primary features are the cluster of resonance lines in the \SIrange{50}{70}{\giga\hertz} range and the $1^-$ line at \SI{118.8}{\giga\hertz}; since the sign of the circular polarization is reversed when transitioning from frequencies below the resonance frequency of a given line and above it, there are nulls due to the interactions between the $1^-$ line and the other resonance lines. Furthermore, the polarized signal is attenuated due to water vapor at \SI{183}{\giga\hertz} and to a lesser extent at \SI{22}{\giga\hertz}.

\begin{figure}
\includegraphics[width=\columnwidth]{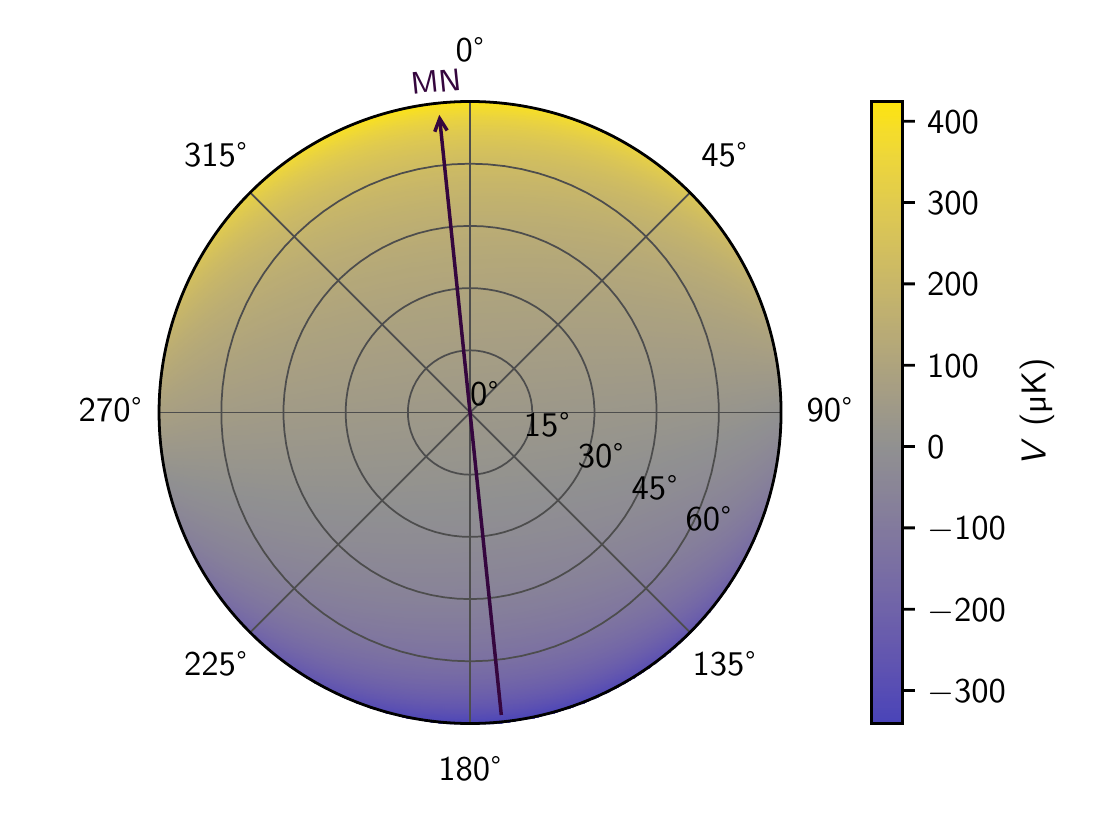}
\caption{Simulated azimuth and zenith angle dependence of the atmospheric Stokes $V$ signal at the CLASS observing site for the CLASS Q-band telescope. Azimuth is shown for a full \SI{360}{\degree}, and zenith angle is shown for \SIrange{0}{75}{\degree}. The \colorindicator{xkcd:darkpurple}{dark purple} arrow indicates magnetic north.}
\label{fig:sky-plot}
\end{figure}

\begin{figure}
\includegraphics[width=\columnwidth]{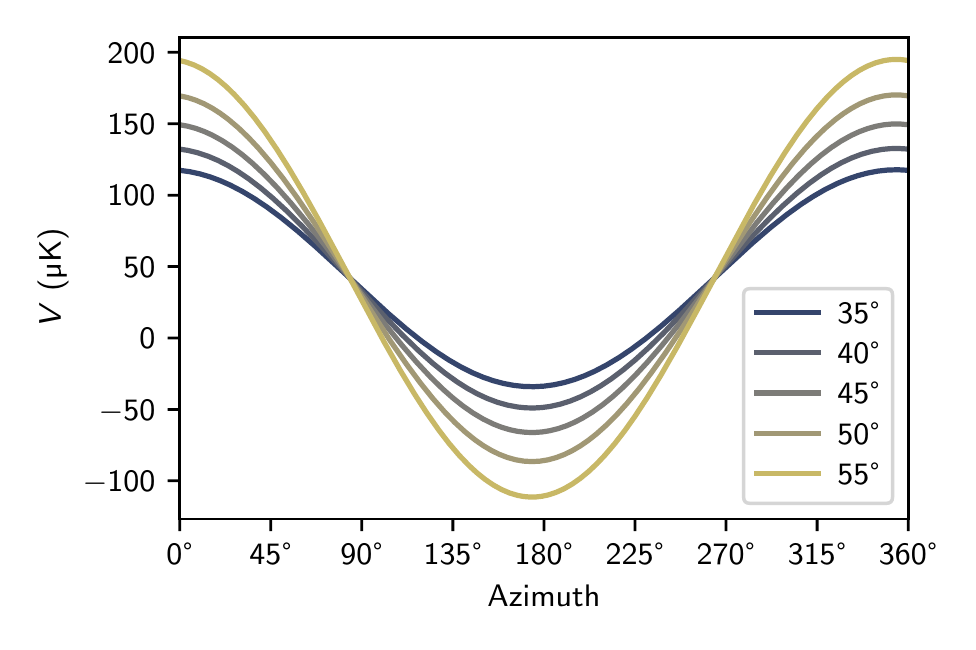}
\caption{Simulated azimuth profiles of the atmospheric $V$ signal at the CLASS observing site for CLASS Q-band telescope are shown for different zenith angles.}
\label{fig:azimuth-dependence}
\end{figure}

Next, effects from \SIrange{32.3}{43.7}{\giga\hertz}, the frequency band of the CLASS Q-band telescope \citep{Appel2019}, are considered. A sky plot is shown in Figure \ref{fig:sky-plot}, which shows both the azimuth and zenith angle dependence of the circularly polarized atmospheric signal. A detailed view of the azimuth dependence is shown in Figure \ref{fig:azimuth-dependence}. The magnitude of the circularly polarized signal is strongest when the azimuth is aligned with the magnetic declination angle, as expected, and is also stronger in the north than the south, since the line-of-sight is better aligned with the magnetic field in that direction, as the magnetic field vector points above the horizon to the north. Since a larger air mass is observed closer to the horizon, the signal is also stronger further from zenith. The effect of the \SI{1.5}{\degree} full width at half maximum beams of the CLASS Q-band telescope on the zenith angle and azimuth dependence was evaluated and found to be negligible, so the effect was not considered in the remainder of the analysis.

For rapid evaluation of the atmosphere model, a function of the form
\begin{equation}
V = a\cdot\tan(b\cdot\phi)\cdot\cos(\psi - c) + d
\label{eq:model-fit}
\end{equation}
can be precisely fit to the simulations, where $\phi$ is the zenith angle and $\psi$ is the azimuth angle, with fit parameters shown in Table \ref{tab:fit}; zenith angles from \SIrange{30}{60}{\degree} were used in the fit. With $a$ and $d$ specified in \si{\kelvin}, $V$ is also in \si{\kelvin}; $b$ is a dimensionless scale factor, and $\psi$ and $c$ are angles. The average fit residual is \SI{\sim200}{\nano\kelvin}. Parameter $c$, the azimuth offset, corresponds exactly to the magnetic declination. The simulation code used to produce these results has been published \citep{Petroff2019}.

\begin{deluxetable}{cl}
\tablecaption{Atmosphere simulation fit parameters at the CLASS observing site for the CLASS Q-band telescope\label{tab:fit}}
\tablehead{
\colhead{Parameter} &
\colhead{Value}
}
\startdata
$a$ & \SI{1.106e-04}{\kelvin} \\
$b$ & \num{9.848e-01} \\
$c$ & \multicolumn{1}{c}{\SI{-5.9}{\degree}} \\ % This is currently fudged, since magnetic field direction in the simulation needs to be fixed
$d$ & \SI{4.185e-05}{\kelvin} \\
\enddata
\tablecomments{When using $b$ with equation (\ref{eq:model-fit}), $\phi$ should be in radians.}
\end{deluxetable}

\section{Comparison with observations}
\label{sec:observations}

From the CLASS Era~1 survey, nighttime data recorded during the period from 2016 September through 2018 February are used for the present analysis; some nights are excluded due to particularly poor weather conditions or due to operational difficulties, such as interruptions to power or cryogenic systems. For each nightly observation, the telescope was scanned with the boresight center pointing at a \SI{45}{\degree} zenith angle through an azimuthal range of \SI{\pm 360}{\degree} at a rate of \SI{1}{\degree\per\second}. For the entirety of each night, the boresight rotation angle of the telescope relative to the horizon remained fixed at \SIlist[list-final-separator={, or }]{-45;-30;-15;0;+15;+30;+45}{\degree}; the boresight rotation angle was changed daily such that each angle was observed on a weekly basis. Boresight rotation combined with individual feedhorn pointing offsets provides access to a range of zenith angles from \SIrange{35}{55}{\degree} with this scanning strategy.

Of the 28 detector pairs in the Q-band telescope that were operational during the Era~1 survey \citep{Appel2019}, 25 are used for the present analysis, with the remaining three conservatively rejected due to atypical noise properties. Anomalous artifacts found in detector timestreams are excised, along with windows surrounding them chosen such that any filtering operations in the timestream processing pipeline do not convolve the artifacts with surrounding data. The Stokes $V$ signal is extracted from pair-differenced detector timestreams via demodulation with the VPM polarization transfer function \citep[K. Harrington et al. 2020, in preparation;][]{Chuss2012}, where each pair corresponds to the two detectors with orthogonal linear polarization sensitivity in each feedhorn. These resulting $V$ timestreams are further checked for stable noise properties, with variance cuts used to eliminate data with abnormally high noise. This data processing pipeline will be described in detail by J. Eimer et al. (2020, in preparation). After all data cuts, 47927.6 detector pair-hours of data remain. Daytime data were excluded from the present analysis, as additional pipeline developments are required to properly handle their reduced stability and artifacts; however, this does not preclude future use of these data. Furthermore, sun avoidance maneuvers alter the telescope's scan strategy during parts of the day and prevent coverage of the full azimuth range, making these data less suitable for the present analysis.

\begin{figure}
\includegraphics[width=\columnwidth]{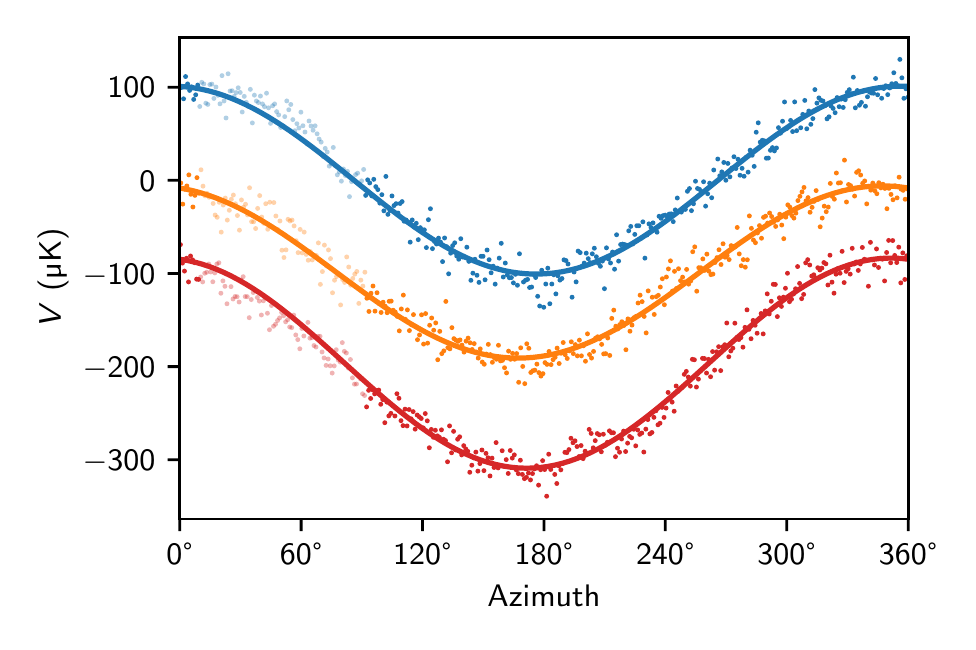}
\caption{Example binned azimuth profiles are shown along with sinusoidal best fit lines for three combinations of detector pairs and boresight rotation angles. The data are plotted with arbitrary amplitude offsets; error bars are not shown. The lighter data points indicate the region excluded from the sinusoidal fits due to the ground elevation angle cut. The profile in \colorindicator{tab:blue}{blue} is from a zenith angle of \SI{43.9}{\degree} and a boresight rotation angle of \SI{-45}{\degree}, the profile in \colorindicator{tab:orange}{orange} is from a zenith angle of \SI{46.7}{\degree} and a boresight rotation angle of \SI{0}{\degree}, and the profile in \colorindicator{tab:red}{red} is from a zenith angle of \SI{52.8}{\degree} and a boresight rotation angle of \SI{+45}{\degree}.}
\label{fig:example-profiles}
\end{figure}

To confirm that an observed circularly polarized signal is due to atmospheric Zeeman emission, three properties should be satisfied: the azimuth angles of maximum and minimum signal should align with the magnetic declination, the signal should show an appropriate zenith angle dependence, and the signal should have approximately the correct amplitude. To evaluate the observed signal, detector data were divided by detector pair and boresight angle, processed as previously described, binned by azimuth with inverse variance weighting, and fit with a sinusoidal profile; variance was evaluated for timestream segments after splitting the demodulated timestreams into sweeps of constant direction scanning in azimuth. As ground pickup contamination, likely due to $T \to V$ leakage, was visible in binned azimuth profiles for some combinations of detector pairs and boresight rotation angles, the azimuth range where the ground elevation angle is above \SI{6}{\degree}, i.e., where the peak of Cerro Toco is, was masked before fitting; this corresponds to \SIrange{10}{92}{\degree} azimuth. A covariance threshold was also used to remove poor fits. As each azimuth sweep is mean-subtracted, only the peak-to-peak amplitude of the azimuth dependent signal is measured, not its absolute offset. Example binned azimuth profiles and sinusoidal fits are shown in Figure \ref{fig:example-profiles}.

\begin{figure*}
\centering
\includegraphics[width=2\columnwidth]{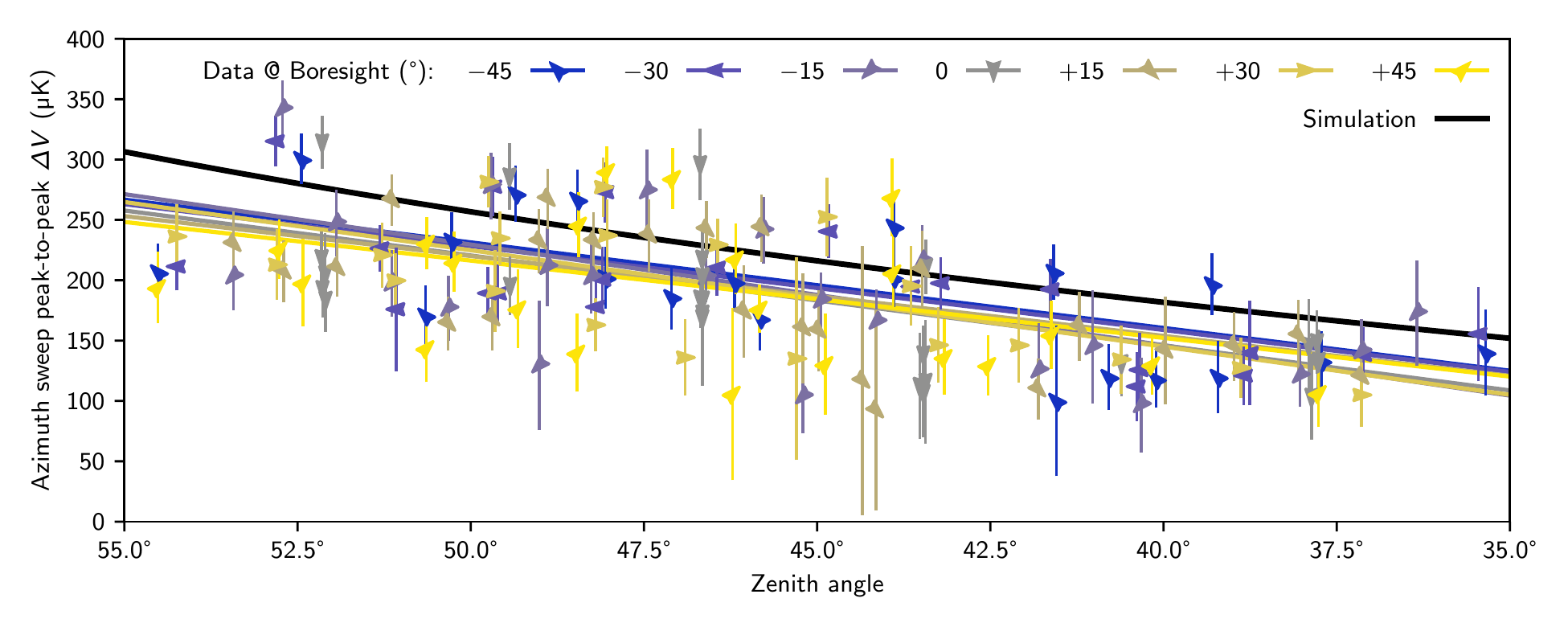}
\caption{Zenith angle dependence of peak-to-peak observed $\Delta{}V$ signal (shown in color) compared to simulation (shown in black). The data are split by both feedhorn and boresight rotation angle, with an individual point plotted for each combination. Error bars show one standard deviation calculated on the residual after subtracting the sinusoidal best fits from the binned azimuth profiles. Linear best fit lines are also shown for each boresight rotation angle; the reduced $\chi^2$ for these fits range from 1.8 to 3.4, with an average reduced $\chi^2$ of 2.5.}
\label{fig:comparison}
\end{figure*}

The fits result in a signal maximum aligned with an azimuth angle of \SI{-5.5 \pm 0.6}{\degree}, which is consistent with the expected magnetic declination, the angle between magnetic north and true north, of \SI{-5.9}{\degree}; this error was calculated using bootstrapping after performing an additional data cut based on the residual after subtracting the sinusoidal best fits from the binned azimuth profiles. The other mechanism by which the Earth's magnetic field is expected to affect the detector data is through magnetic pickup in the detectors and Superconducting Quantum Interference Device (SQUID) multiplexers and amplifiers. As this pickup is not modulated by the VPM and should therefore not be visible in the demodulated timestreams, this alone is strong evidence of a detection of a signal of atmospheric origin. Next, the zenith angle dependence of the peak-to-peak signal in the data was evaluated and compared to the simulation, the results of which are shown in Figure \ref{fig:comparison}. A data point is shown for each combination of detector pairs and boresight rotation angles; zenith angle differences are due to individual feedhorn pointing offsets, which change relative to the boresight center pointing at different boresight rotation angles. An expected zenith angle dependence is seen, giving further evidence that the observed signal is of atmospheric origin. The excess scatter is thought to be due to uncertainties in the preliminary detector relative efficiency calibrations used in the analysis combined with potential bandpass mismatches between detectors.

As can be seen in Figure \ref{fig:comparison}, the measured amplitude is consistent between different boresight rotation angles and between different detector pairs, providing a check against systematic errors. Additionally, data were split by date of observation to check for changes over time; these splits were also found to be consistent. While there are expected to be slight changes over time to the atmospheric signal due to evolution of the geomagnetic field, these changes are much smaller than the error of the measurements; EMM2017 predicts the yearly change in field strength to be on the sub-percent level and the yearly change in magnetic declination to be approximately \SI{-0.2}{\degree}.

A ground loop between the VPM control electronics and the detector readout electronics would in theory be able to introduce modulated signal into the SQUID magnetic pickup. However, detector data collected while the VPM was running and the receiver window was covered with a metal plate to block optical signals was inspected and found to contain no modulated signal, ruling out this possibility.

To further exclude the possibility of the detected signal being due to magnetic pickup, the physical geometry of the telescope's receiver can be considered. The detectors and SQUIDs are all in the same plane relative to one another despite having different optical pointing offsets, i.e., the detectors all have the same physical orientation relative to the magnetic field. Thus, the peak-to-peak amplitude of their magnetic pickup---which should not be in the demodulated timestreams to start with---would only be dependent on the zenith angle of the telescope's boresight pointing. Since all observations were taken with a boresight pointing zenith angle of \SI{45}{\degree}, SQUID magnetic pickup would not result in a zenith angle dependence, contrary to the detected signal. Additionally, changing the boresight rotation angle alters the physical orientation of the SQUIDs relative to the magnetic field and thus yields a boresight rotation angle dependence to the magnetic pickup, again contrary to the detected signal.

\begin{figure}
\includegraphics[width=\columnwidth]{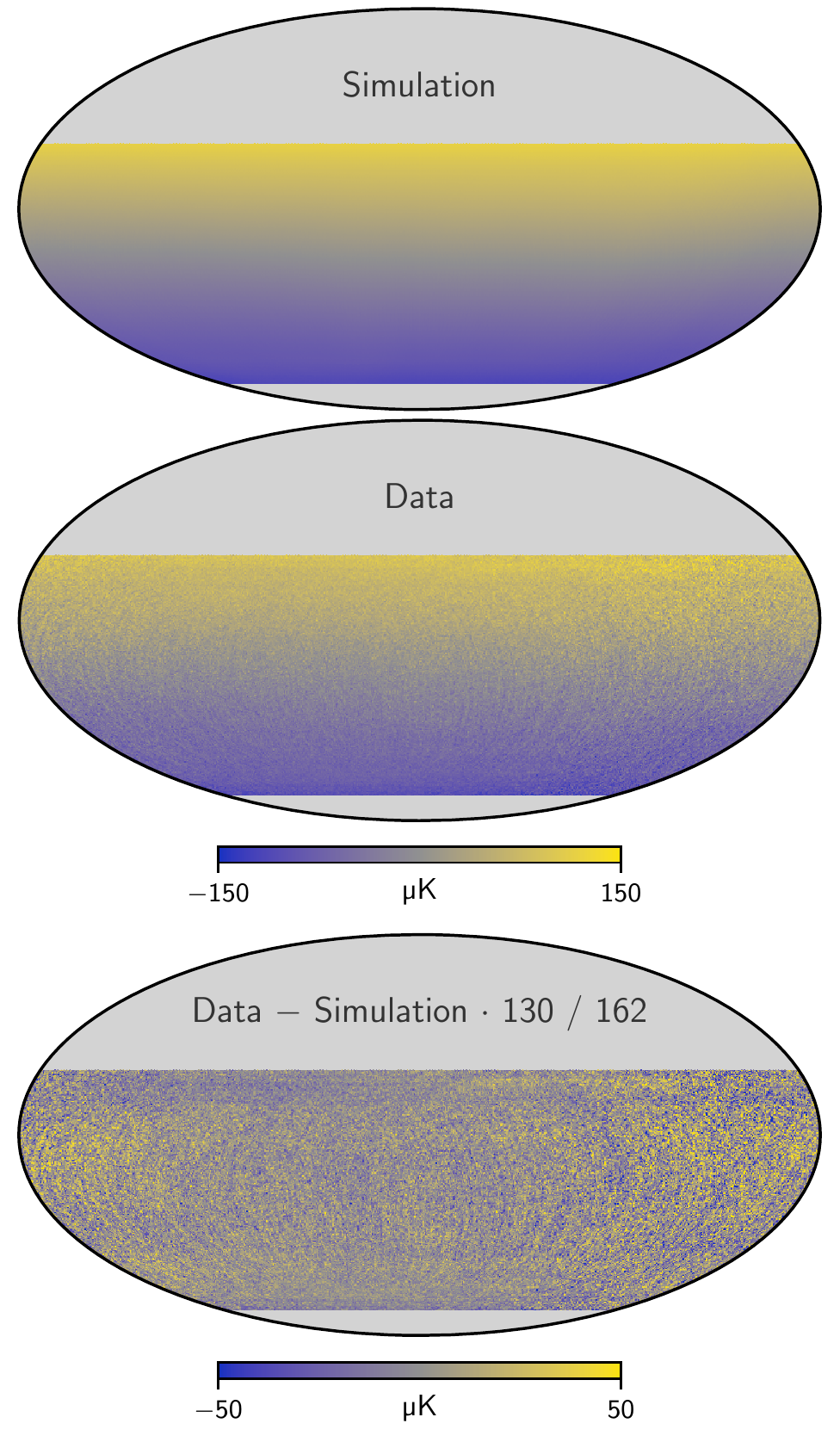}
\caption{Stokes $V$ signal mapped onto the projected celestial sky using pointings from CLASS Q-band telescope observations. The top plot uses a simulated atmospheric signal, the middle plot uses data, and the bottom plot shows the difference between the two maps, with the simulation data scaled by the amplitude difference of the dipoles fitted to the two previous maps. The color scale is identical for the top two maps and different for the bottom map. Note that this data map was produced with a different analysis pipeline than the maps presented in \citet{Padilla2019}.}
\label{fig:sim-map}
\end{figure}

Using equation (\ref{eq:model-fit}) and the fit parameters in Table \ref{tab:fit}, a simulated map was created using pointings from the aforementioned observations, which is shown in Figure \ref{fig:sim-map} along with a map of CLASS $V$ data and a map showing residuals. To create the data map, the $V$ signal was accumulated into pixels using inverse variance weighting of data, using the same data processing and variance calculation procedures used for fitting the sinusoidal profiles. For the simulated map, azimuth sweeps were mean subtracted and weighted to match the processing of the data. Note that the overall dipole pattern is visible in both maps; the dipole amplitude is \SI{162}{\micro\kelvin} in the simulated map and \SI{130}{\micro\kelvin} in the data map, with a direction difference of \SI{0.7}{\degree}. The dipole amplitude ratio between the simulated map and the data map was used to scale the simulated map before subtracting it from the data map to produce the residual map. If the azimuth sweeps are not mean subtracted when creating the simulated map, the dipole changes in amplitude by \SI{+6}{\micro\kelvin} and in direction by \SI{0.4}{\degree}, demonstrating that the mean subtraction only has a minor effect on the signal. Areas of higher noise are due to uneven sky coverage.

The measured amplitude in both the sinusoidal profiles and map are in reasonable agreement to that predicted by the simulations. The remaining ${\sim}20\%$ discrepancy is likely due to some combination of calibration errors in the detector data and shortcomings of the simulation. Although the detector calibration was done using Stokes $I$, the demodulated linear polarization of the Crab Nebula matches previous observations \citep{Xu2019}, so it is unlikely that a discrepancy of the magnitude observed between observation and simulation is due to an error in either the VPM polarization transfer function calibration or detector calibration. The detector bandpass uncertainty is also low enough that a bandpass error cannot fully explain the discrepancy \citep{Appel2019}. This leaves a shortcoming of the simulation as the most likely source of the discrepancy. The Van~Vleck--Weisskopf line shape is known to be inaccurate far from the resonance lines \citep{Hill1986}, so it may be reasonable to attribute the difference to this inaccuracy, since we are observing at frequencies far from the resonance lines, and the majority of the observed signal is from the lower atmosphere, where the line profile used in the simulations reduces to a Van~Vleck--Weisskopf line shape. Furthermore, the line mixing and pressure broadening parameters used in the simulation were also all measured near the resonance lines. As Zeeman splitting of molecular oxygen resonance lines in the presence of Earth's magnetic field is the only theorized source of circularly polarized atmospheric emission, we conclude that this is the source of the detected signal.

\section{Conclusion}
\label{sec:conclusion}

Expanding on prior models and utilizing recent spectroscopic data, a model for circularly polarized atmospheric emission from Zeeman splitting of molecular oxygen was presented. This model was then used to simulate the atmosphere of the CLASS observing site at the frequencies of the CLASS Q-band telescope. An analysis of circular polarization timestreams observed by the CLASS Q-band telescope, utilizing the Stokes $V$ measurement capability of a VPM, was then compared to the simulations and shown to be a strong detection of this atmospheric emission; this is believed to be the first such detection of circular polarization in a frequency band used for ground-based CMB observations. The amplitude of the signal differed by ${\sim}20\%$ between the simulations and observations but is still in good agreement. As the atmospheric signal is orders of magnitude larger than any theoretical cosmological or astrophysical signal, its subtraction is necessary for setting improved upper limits on said signals and providing more rigorous observational tests. Although this signal can be adequately removed empirically, such as by fitting a dipole sky pattern or by creating templates that describe the signal as a function of azimuth, modeling the signal helps us verify our understanding of it. After the subtraction of the atmospheric signal, Stokes $V$ serves primarily as a null channel for CLASS; as no residual signal is expected, a $VV$ angular power spectrum that is not consistent with zero would suggest the presence of either an unmitigated systematic error or a non-standard cosmological signal.

%\acknowledgments
% Needed because \acknowledgments is buggy; copied from aastex63.cls (without box)
\vskip 5.8mm plus 1mm minus 1mm
\vskip1sp
\section*{Acknowledgments}
\vskip4pt

We acknowledge the National Science Foundation Division of Astronomical Sciences for their support of CLASS under Grant Numbers 0959349, 1429236, 1636634, and 1654494. The CLASS project employs detector technology developed in collaboration between JHU and Goddard Space Flight Center under several previous and ongoing NASA grants. Detector development work at JHU was funded by NASA grant number NNX14AB76A. K. Harrington was supported by NASA Space Technology Research Fellowship grant number NX14AM49H. We thank the anonymous reviewer for providing astute comments and suggestions that helped improve this manuscript. We acknowledge scientific and engineering contributions from Max Abitbol, Mario Aguilar, Fletcher Boone, David Carcamo, Francisco Espinoza, Saianeesh Haridas, Connor Henley, Yunyang Li, Lindsay Lowry, Isu Ravi, Gary Rhodes, Daniel Swartz, Bingie Wang, Qinan Wang, Tiffany Wei, and Ziang Yan. We thank Mar\'ia Jos\'e Amaral, Chantal Boisvert, William Deysher, and Miguel Angel D\'iaz for logistical support. We acknowledge productive collaboration with Dean Carpenter and the JHU Physical Sciences Machine Shop team. Part of this research project was conducted using computational resources at the Maryland Advanced Research Computing Center (MARCC). We further acknowledge the very generous support of Jim and Heather Murren (JHU A\&S '88), Matthew Polk (JHU A\&S Physics BS '71), David Nicholson, and Michael Bloomberg (JHU Engineering '64). CLASS is located in the Parque Astron\'omico Atacama in northern Chile under the auspices of the Comisi\'on Nacional de Investigaci\'on Cient\'ifica y Tecnol\'ogica de Chile (CONICYT). R. D\"unner and P. Flux\'a thank CONICYT for grant BASAL CATA AFB-170002. R. Reeves acknowledges partial support from CATA, BASAL grant AFB-170002, and CONICYT-FONDECYT, through grant 1181620.

\software{Astropy \citep{Astropy2013, Astropy2018}, NumPy \citep{vanderWalt2011}, SciPy \citep{Scipy2019}, Numba \citep{Lam2015}, MSISE-00 \citep{Hirsch2019}, Healpy \citep{Zonca2019, Gorski2005}, Matplotlib \citep{Hunter2007}}

\bibliography{paper.bib}

\end{document}